# Brand Attitude in Social Networks: The Role of eWoM

Azita Pourkabirian [1], Melika Habibian [2], Azadeh Pourkabirian [3, *]

*[1]Department of Public Administration, Isfahan Branch, Azad University, Isfahan, Iran*
*[2]Department of Computer Engineering, Imam Khomeini International University, Qazvin, Iran*
*[3]Faculty of Computer and Information Technology Engineering, Qazvin Branch, Islamic Azad University, Qazvin, Iran*

**Abstract**

This study aims to examine the impact of electronic word-of-mouth (eWoM) marketing on branding attitudes in social networks. Specifically, we investigate the effects of eWoM activities on brand awareness, brand destruction, branding, brand image, and brand competition. To gather data, we conducted a survey among followers of the Vizland shoe page on the Instagram social network. Statistical analysis was performed using SPSS software, and hypotheses were tested using SmartPLS software. The results indicate that eWoM significantly and positively influences branding, brand image, and brand awareness, while it does not have an impact on brand destruction. Additionally, branding and brand destruction are found to play a crucial role in gaining competitive advantage. Therefore, eWoM activities contribute to enhancing the brand relationship.

## 1. Introduction

In today's global economy, service providers are competing to attract and retain customers by offering greater satisfaction and loyalty (Hassan *et al*., 2014). Customers who have a strong relationship with the supplier are recognized as a major source of profit. Therefore, companies must understand their customers' behavior and meet their needs in order to retain them. With the rise in competition and advertising, brands face challenges in capturing consumers' attention, conveying their message effectively, and ultimately convincing them to purchase their products (Minarti & Segoro, 2014). The emergence of computers, the internet, and new technology has provided consumers with access to more information about brands, products, and services. This enables them to make more informed and personalized choices based on their individual circumstances. The internet and computer networks have also transformed communication, creating a cyberspace where individuals with shared interests can interact electronically. Customers and consumers on social networks utilize the available information resources to become advocates for specific brands. As a result, social networks have a significant impact on a brand's success, in addition to traditional communication tools. Customers can either endorse a company's marketers or challenge their claims based on their knowledge (Gupta *et al*., 2016). Word-of-mouth marketing is considered more trustworthy than other techniques, as only 14% of people trust commercial advertisements compared to the 90% who trust recommendations from family, friends, or colleagues (Jalilvand and Ebrahimi, 2011). Electronic word-of-mouth (eWoM) advertising encompasses various media forms and websites that provide easy access to statistics, opinions, and reviews from online consumers (Zhang *et al*., 2010). In cyberspace, users and consumers can contribute their own reviews, comments, and evaluations of products on blogs, discussion forums, review sites, Usenet newsgroups, and social media platforms. Research has shown that word-of-mouth can influence product evaluations (Gruen, 2006) and the perceived credibility and loyalty towards a company's products (Radmehr *et al*., 2011). Therefore, studying brand loyalty and customer behavior through new marketing methods and online advertising in social media is crucial for companies to succeed in today's competitive markets (Ghafourian Shagerdi *et al*., 2017). In recent years, industrial organizations have recognized that the company's name is a valuable asset. The higher the brand's value in the minds of consumers, the more benefits the company can gain from consumer support (Sharifi and Ansari, 2015). Consequently, building strong brands has become a marketing priority due to the significant advantages they bring to organizations

[*] Corresponding author. Tel: +91-80-22082823; 9880148294

E-mail: arun@jncasr.ac.in

(Rasoulzadeh and Pourkabirian, 2020). Strong brands establish a distinct identity for a company in the market. Over the past decades, researchers and practitioners in the field of goods and services have paid close attention to branding. Brands hold a value that extends beyond traditional financial assets created through professional activities.

Cyberspace offers marketers new opportunities to gain and retain customers, create a brand image, and effectively communicate with customers (Litvin *et al*., 2008). People have a significant influence on customers' evaluations and purchasing decisions in online environments. Research has shown that word-of-mouth (WOM) communication is more effective than other sources such as top news recommendations or advertisements in digital newspapers because it provides reliable information (Jalilvand and Samiee, 2012). Electronic word-of-mouth (eWoM) advertising refers to informal communications among online users about the usage or features of goods and has become an important platform for consumer feedback (Sölvell, 2015). It is more efficient than offline WOM communication. Therefore, it is necessary to study the impact of eWoM on branding, business competition, and the macro-level business environment. Previous studies have focused on the impact of eWoM on customer purchase intention, loyalty, and brand image. However, this research aims to investigate the impact of eWoM on branding and brand destruction. Specifically, we will examine the effects of eWoM on brand image, brand advertisement, and their individual impacts. The results of this study will provide practical insights for business owners on employing electronic marketing in a competitive market and determining its impact on branding. The following sections will discuss the background of the research, research methods, results, and suggestions.

## 2. Literature Review

There has been extensive research conducted on the effects of eWoM in various business environments. For example, Ismagilova *et al*. (2017) addressed several important issues related to eWoM advertising in marketing. Abubakar *et al*. (2017) explored the relationship between eWoM and trust intention, focusing on gender differences. Their findings indicated a significant influence of two key eWoM advertising factors on decision reviewing and destination trust within the tourism industry. Liang *et al*. (2018) investigated the impact of eWoM advertising on customer repurchase intention, collecting data from 395 respondents in Canada and the United States. Hudson *et al*. (2015) examined the effect of eWoM on customers' purchasing goals. Murtiasih & Siringoringo (2013) explored the impact of eWoM on brand equity in the automotive industry in Indonesia. Liang *et al*. (2018) further investigated the influence of content type on personal participation and social media promotion using the total addressable market (TAM) and the elaboration likelihood model (ELM). They found that individual participation and WOM effect on social media were influenced by content type. Additionally, Hassan *et a*l. (2014) explored the influence of eWoM participation on customer loyalty and online shopping patterns.

## 3. Methodology

This research can be classified as descriptive-cohesion as it aims to understand the effective factors on branding in social media. Additionally, it can be considered correlational research as it examines the effectiveness of variables and their correlation with each other. The purpose of this research is analytical, aiming to provide insights into the field. The research results can serve as a valuable model for understanding the impact of eWoM in various business environments.

In this study, the non-probability snowball sampling method was used to overcome the challenge of accessing the total target community. An online questionnaire was created and made available to multiple users on social media, who were then asked to share the questionnaire with other users. This sampling method ensured that each member of the community had an equal chance of being selected for the sample. For a community with a population size of N and a sample size of n, the probability of selecting each individual in this sampling method is n/N. Simple random sampling is typically used when the population structure is coherent. In this method, the sample is randomly selected from the statistical population without any previous order or planning. In this research, the statistical population consisted of approximately 10,000 users who followed the Vizland shoe store page on Instagram. To estimate the sample size, Morgan's table can be used assuming no error rate or community variance. According to Morgan's table, the minimum sample size required for this research was approximately 384 individuals.

In order to collect data for the main part of the research and examine the research variables, an online questionnaire was used. This questionnaire was made available to followers of the Vizland shoe store page on Instagram. Before being used, the questionnaire was reviewed and evaluated by experts in the field. The questionnaire consisted of two parts. The first part collected demographic information, while the second part consisted of 28 closed-ended questions using a Likert scale to measure the research variables. Each answer on the Likert scale had four options: completely agree, agree, disagree, and completely disagree, each with a corresponding score from one to five. This allowed for the transformation of qualitative and non-parametric information into quantitative and numerical values, which were used as criteria in the analysis. To assess the homogeneity of the reflective measurement model, the factor loading values of observed variables were evaluated. It was found that each observed variable corresponding to the latent variable had a factor loading value of at least 0.4. Additionally, significance levels of 90%, 95%, and 99% were used to compare the upper limit of the minimum statistic with the minimum t-statistic values of 1.64, 1.96, and 2.85, respectively. Various methods of descriptive and inferential statistics and hypothesis testing were utilized to analyze the data. The collected data from the questionnaires were analyzed using SPSS and SmartPLS software in both the descriptive and inferential sections. In the descriptive section, SPSS software was used to improve operations related to the statistical sample's demographic information. In the inferential section, structural equation modeling (SEM) was employed to answer the research hypotheses. This method is suitable for measuring the overall impact of structures on each other, as it takes into account the measurement errors of each variable. One of the advantages of using SmartPLS software is its ability to analyze data with low volume. Additionally, this software does not require data normalization.

## 4. Results

The findings of the descriptive analysis revealed that out of the total respondents, 310 were male and 74 were female. The frequency and percentage values of the sample distribution indicated that the male group was four times larger than the female group. In terms of age, 45.1% of the respondents were under 30 years old, 36.7% were between 31 and 40 years old, 14.6% were between 41 and 50 years old, and 3.6% were above 50 years old. Regarding education level, 20.6% of the respondents had high school diplomas or less, 14.3% had associate degrees, 46.3% had bachelor's degrees, and 18.8% had degrees higher than a bachelor's degree. With respect to work experience, the sample distribution results showed that 39.8% of individuals had less than 3 years of experience, 34.4% had between 3 and 5 years, 14.8% had between 5 and 10 years, and 11% had more than 10 years of experience.

The results of the reliability analysis indicate that the items in the research measurement model have factor load values above 0.4, demonstrating homogeneity and acceptability. This means that there is no need to remove any items. Additionally, the t-statistical values for all items exceed 2.58, indicating a significant relationship between the items and their latent variable at a confidence level of 99%. Composite reliability and Cronbach's alpha values for all latent variables are above 0.7, confirming the reliability of the measuring instruments. The validity of the measuring instruments was also assessed. The extracted variance values for all latent variables were found to be greater than 0.5, demonstrating convergent validity. Furthermore, the discriminant validity of the measurement model was evaluated by comparing the factor load of each variable divided by its latent variable to the factor load of the same observed variable divided by other latent variables. The results showed that the former is greater than the latter, confirming differential validity. The Fornell-Larker criterion was used to verify the divergent validity of the measurement model (Table 1).

Table 1

Fornell-Larcker criterion

| | EWOM | Brand Awareness | Branding | Brand Destruction | Brand Image | Brand Competition |
|---|---|---|---|---|---|---|
| Brand Competition | 0.301 | 0.393 | 0.441 | 0.709 | 0.305 | 0.715 |
| Brand Image | 0.645 | 0.689 | 0.594 | 0.253 | 0.835 | - |
| Brand Destruction | 0.342 | 0.394 | 0.393 | 0.716 | - | - |
| Branding | 0.554 | 0.557 | 0.702 | - | - | - |
| Brand Awareness | 0.701 | 0.718 | - | - | - | - |
| EWOM | 0.764 | - | - | - | - | - |

The following are some descriptive parameters of variables including mean, standard deviation, skewness, and kurtosis in Table 2. For instance, the average of electronic word-of-mouth advertising in this study is 3.72 which indicates that most of the data related to this variable are centered around this point.

Table 2

Central, Dispersion and Distribution Form indexes

| Index | | EWOM | Brand Awareness | Branding | Brand Destruction | Brand Image | Brand Competition |
|---|---|---|---|---|---|---|---|
| Central | Mean | 3.723 | 3.973 | 3.398 | 3.236 | 3.376 | 3.213 |
| Dispersion | Standard Deviation | 0.689 | 0.543 | 0.324 | 0.546 | 0.786 | 0.685 |
| | Variance | 0.825 | 0.579 | 0.479 | 0.276 | 0.495 | 0.494 |
| Distribution Form | Skewness | 0.45 | 0.564 | 0.324 | 0.123 | 0.143 | 0.129 |
| | kurtosis | 0.123 | 0.543 | 0.132 | 0.154 | 0.156 | 0.147 |

The other descriptive parameters in the community are dispersion parameters. One of the most important dispersion parameters is the standard deviation. The higher the standard deviation of a statistical distribution is, the more scattered the data is. Among the variables of this study, the brand image has the highest distribution with a value of 0.786. The observed skewness value for the studied variables is in the range (-2, 2). This means that in terms of skew, the research variables are normal and their distribution is symmetrical. The kurtosis value of the variables is also in the range (-2, 2). It indicates that the distribution of variables has a normal kurtosis.

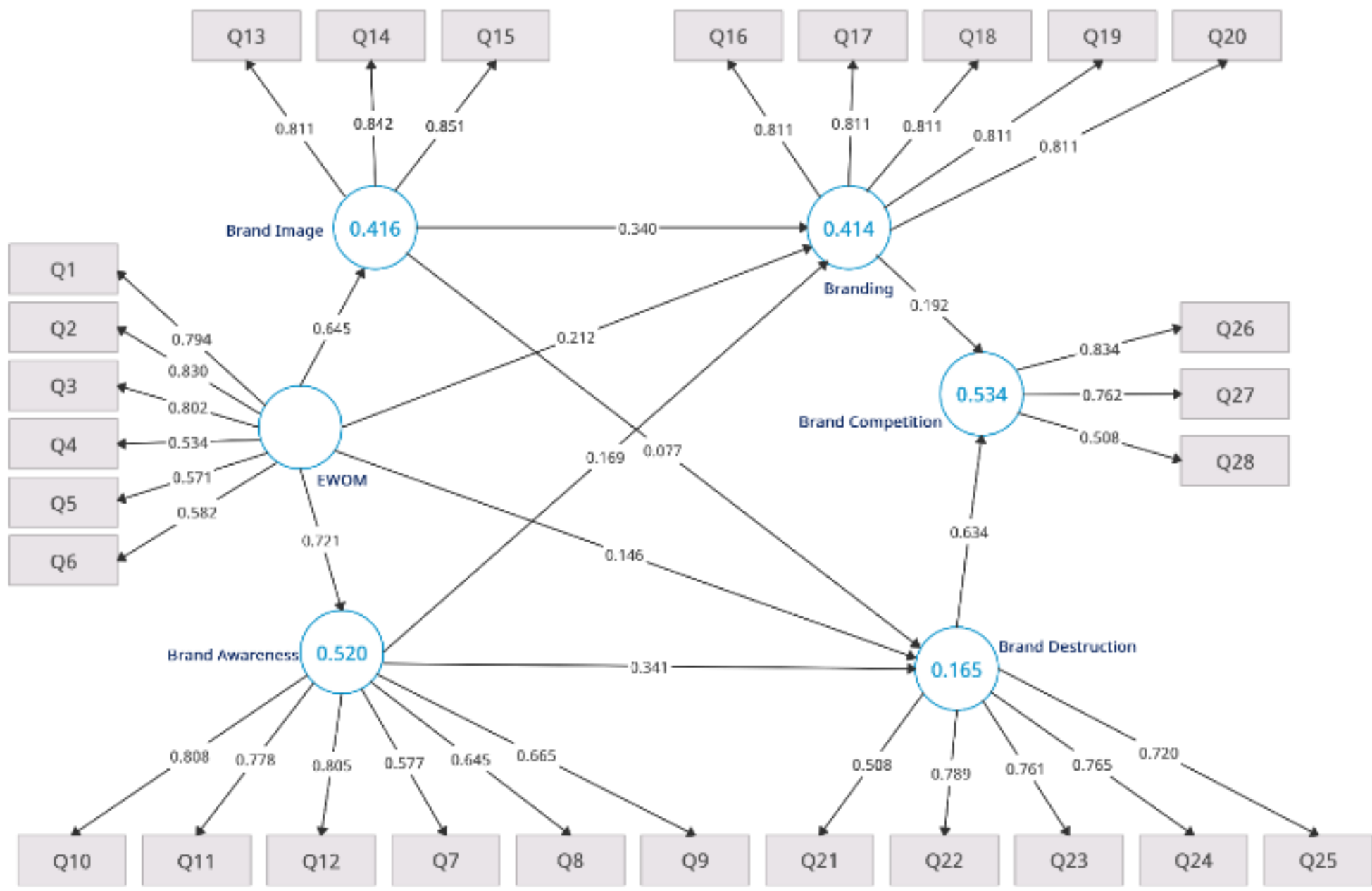

Fig. 1. Path Coefficient.

The value of the path coefficient is in the range (-1, 1). The higher is this positive value, the greater is the effect of the independent variable on the dependent variable. One of the indicators of confirming connections in the structural model is the significance of path coefficients. The significance of the path coefficients is the complement of the size and the sign of the beta coefficient in the model. If the upper limit value of the minimum statistic is considered at the reliable level, that relation or hypothesis is confirmed. At the significance level of 90%, 59%, and 99%, this value is compared with the minimum t-statistic of 1.64, 1.96, and 2.58, respectively. Fig. 2 depicts the reliability of path coefficients for this study.

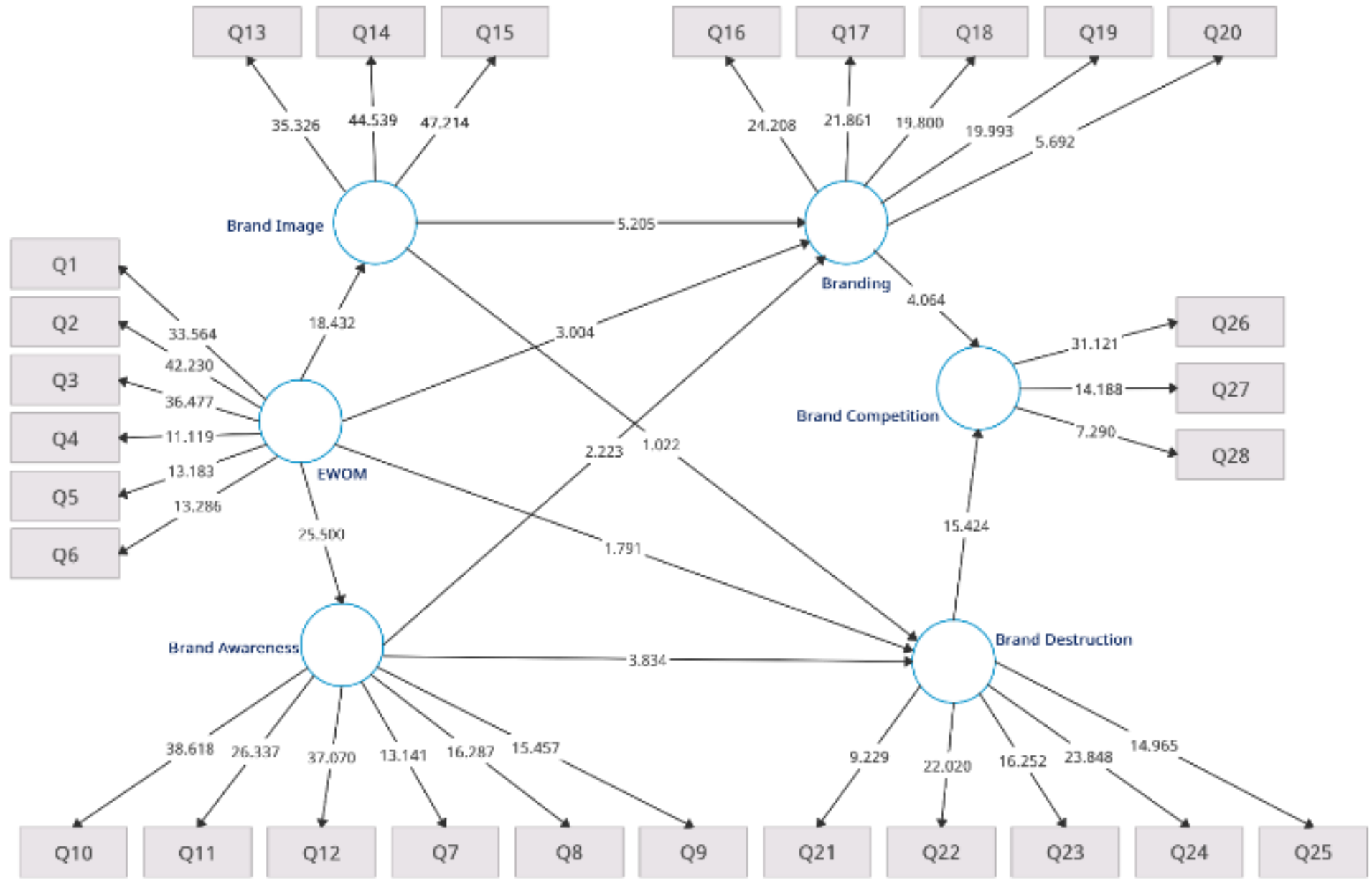

Fig. 2. path coefficients meaningfulness

Moreover, to examine the ability of the structural model in predicting, the prediction performance of the model is presented in Table 3 from weak to strong.

Table 3

Prediction Performance of the Model

| | $Q^2$(=1-SSE/SSO) |
|---|---|
| Brand Awareness | 0.248 |
| Branding | 0.248 |
| Brand Destruction | 0.248 |
| Brand Image | 0.151 |
| Brand Competition | 0.151 |

The results of the analysis show that the predictive power of the model is strong. Generally, models that are examined with variance-based methods by variance-based software such as Smart PLS, have no general index for overviewing models. Therefore, in this study, we use an indicator called GOF which was proposed by Tenenhaus *et al*. (2005) to measure the whole model. This indicator considers both structural and measurement models and tests their quality. It should be noted that Wetzels *et al*. (2009) measured three values of 0.01, 0.25, and 0.36, respectively, and introduced them as weak, medium, and strong values for GOF.

$$\text{GOF} = \sqrt{\overline{communalities} \times \overline{R^2}}$$

The result of fitting the general model is presented in Table 4. Since the obtained value of GOF in this study

(0.476) is higher than Wetzels's proposed value (0.36), therefore, the fitting of the general model is approved.

Table 4

General Model Fitting

| | GOF | √ Shared values | Shared values | √ Coefficient of Determination | Coefficient of Determination |
|---|---|---|---|---|---|
| EWOM | | | 0.585 | | - |
| Brand Awareness | | | 0.516 | | 0.520 |
| Branding | 0.476 | 0.744 | 0.501 | 0.640 | 0.414 |
| Brand Destruction | | | 0.513 | | 0.165 |
| Brand Image | | | 0.697 | | 0.416 |
| Brand Competition | | | 0.511 | | 0.534 |

## 5. Discussion and Conclusions

In today's competitive business environment, it is crucial for managers and capitalists to utilize effective marketing strategies to attract and retain customers. One such strategy is electronic word-of-mouth (eWoM) advertising, which has proven to be highly effective. Simply providing customers with specialized information about a product or service is no longer sufficient, as customers seek information from reliable sources. With the increasing influence of social media on traditional marketing, corporate executives and business owners can leverage this platform to strengthen their competitive advantage. This study aims to evaluate the impact of eWoM advertising on branding and brand destruction, ultimately leading to a competitive advantage. The findings of this study showed that eWoM advertising has a positive and significant impact on branding in social media. The test results of this hypothesis indicated that the value of the path coefficient for eWoM advertising on branding is about 0.212 in social media. The t-statistic for this impact is 3.004, which was reported to be significant at the confidence level of 95% (P-Value ≤ 0.05). The second hypothesis evaluated the impact of eWoM advertising on brand destruction in social media. The results of this study indicated that the value of the path coefficient for eWoM advertising on brand destruction is 0.146 in social media. The t-statistic for this impact is 1.791, which wasn't reported to be significant at the confidence level of 95% (P-Value ≥ 0.05). The evaluation results of the third hypothesis showed that word-of-mouth advertising has a positive and significant effect on brand image in social media. These findings suggested that the value of the path coefficient for eWoM advertising on brand image is 0.645 in social media. The t-statistic for this impact is 18.432, which was reported to be significant at the confidence level of 95%. The next hypothesis examined the effect of word-of-mouth advertising on brand awareness in social media. The value of the path coefficient for eWoM advertising on the image was 0.721 in social media. The t-statistic for this impact was 25.500, which was reported to be significant at the confidence level of 95% (P-Value ≤ 0.05). In the next hypothesis, the brand image was studied in social media on branding. The test results of this hypothesis also showed that the value of the path coefficient is 0.34. The t-statistic for this impact is 5.205, which was reported to be significant at the confidence level of 95%. The test results of the sixth hypothesis indicated that the value of the path coefficient for the brand image on brand destruction is equal to 0.077 in social media. The t-statistic for this impact is 1.022, which wasn't reported to be significant at the confidence level of 95% (P-Value ≤ 0.05). It means that brand image does not have a positive and significant effect on brand destruction in social media. Next, the results of testing the brand awareness hypothesis on branding in social media were calculated, which was 0.169. The t-statistic for this impact is 2.23, therefore it was reported to be significant at the confidence level of 95%. These results suggest that brand awareness has a positive and significant effect on branding in social media. Lastly, the eighth hypothesis of brand awareness in social media was studied on brand destruction. The path coefficient was 0.341. The t-statistic for this impact is 3.83, which was reported to be significant at the confidence level of 95%. These results indicate that brand awareness has a positive and significant effect on brand destruction in social media. Finally, in the last two hypotheses, the impact of branding and brand destruction on competition between brands in social media was studied respectively. The results of the studies showed that both assumptions were confirmed. In other words, branding and also brand destruction have a positive and significant effect on brand competition in social media. Furthermore, the study findings demonstrate that eWoM advertising not only positively and significantly impacts branding, brand image, and brand awareness, but it also has no effect on brand destruction. This highlights the importance of utilizing eWoM advertising as a means to achieve a competitive advantage in today's highly competitive business environment, characterized by rapid technological advancements and increased consumer power. In order to succeed in this environment, companies must have a deep understanding of customer expectations and effectively respond to them. eWoM advertising serves as a crucial source of information for customers and can be leveraged to gain a competitive edge. However, it is important for advertising programs to have an economic justification. Advertising agencies within companies often face the challenge of demonstrating the effectiveness of advertisements in order to increase profits. In this regard, advertising managers expect to utilize advertising to not only enhance the company's profitability but also determine its role in overall company management.

## References

Abubakar, A. M., Ilkan, M., Al-Tal, R. M., & Eluwole, K. K. (2017). eWoM, revisit intention, destination trust and gender. Journal of

Ansari, M., Sharifi, M., & Ansari, N. (2015). Identifying and Ranking the Factors, Affecting Advertising Creativity in Iran's TV Commercials. Journal of Business Management, 7(4), 823-840.

Jalilvand, M., and Ebrahimi, A. (2010). The effect of word-of-mouth communication on domestic car purchases, Case Study of Samand Car of Iran Khodro Co. 9(3).

Gruen, T. W., Osmonbekov, T., & Czaplewski, A. J. (2006). eWoM: The impact of customer-to-customer online know-how exchange on customer value and loyalty. Journal of Business Research, 59(4), 449-456.

Gupta, S., Malhotra, N. K., Czinkota, M., & Foroudi, P. (2016). Marketing innovation: A consequence of competitiveness. Journal of Business Research, 69(12), 5671-5681.

Hassan, L. F. A., Jusoh, W. J. W., & Hamid, Z. (2014). Determinant of customer loyalty in Malaysian Takaful Industry. Procedia-Social and Behavioral Sciences, 130, 362-370.

Hudson, S., Roth, M. S., Madden, T. J., & Hudson, R. (2015). The effects of social media on emotions, brand relationship quality, and word of mouth: An empirical study of music festival attendees. Tourism Management, 47, 68-76.

Ismagilova, E., Dwivedi, Y. K., Slade, E., & Williams, M. D. (2017). Electronic word of mouth (eWoM) in the marketing context: A state of the art analysis and future directions. Springer.

Jalilvand, M. R., & Samiei, N. (2012). The effect of electronic word of mouth on brand image and purchase intention. Marketing Intelligence & Planning.

Liang, L. J., Choi, H. C., & Joppe, M. (2018). Understanding repurchase intention of Airbnb consumers: perceived authenticity, electronic word-of-mouth, and price sensitivity. Journal of Travel & Tourism Marketing, 35(1), 73-89.

Litvin, S. W., Goldsmith, R. E., & Pan, B. (2008). Electronic word-of-mouth in hospitality and tourism management. Tourism Management, 29(3), 458-468.

Minarti, S. N., & Segoro, W. (2014). The influence of customer satisfaction, switching cost and trust in a brand on customer loyalty– The survey on a student as IM3 users in Depok, Indonesia. Procedia-Social and Behavioral Sciences, 143, 1015-1019.

Murtiasih, S., & Siringoringo, H. (2013). How word of mouth influences brand equity for automotive products in Indonesia. Procedia-Social and Behavioral Sciences, 81, 40-44.

Radmehr, R., Rezaee Dolat Abadi, H., and Shalikar, M. (2010). Investigation of Electronic Word of Mouth Advertising in Tourism Management. 1st International Conference on Tourism Management and Sustainable Development. Marvdasht Azad University.

Rasoulzadeh, F., & Pourkabirian, A. (2020). The Effect of Electronic Word-of-Mouth on Brand Destruction and Branding in Social Networks. 2nd International Conference on Information Technology, Computer and Telecommunication. Tehran. Iran.

Ghafourian Shagerdi, A., Daneshmand, B., & Behboodi, O. (2017). The Impact of Social Networks Marketing toward Purchase Intention and Brand Loyalty. New Marketing Research Journal, 7(3), 175-190. 26(3).

Sölvell, Ö. (2015). The Competitive Advantage of Nations 25 years– opening up new perspectives on competitiveness. Competitiveness Review.

Tenenhaus, M., Vinzi, V. E., Chatelin, Y. M., & Lauro, C. (2005). PLS path modeling. Computational statistics & data analysis, 48(1), 159-205.

Wetzels, M., Odekkerken-Schroder, G. & Van Oppen, C. (2009). Using PLS path modeling for assessing hierarchical construct models: Guidelines and empirical illustration, MIS Quarterly, 33(1): 177.

Zhang, J. Q., Craciun, G., & Shin, D. (2010). When does electronic word-of-mouth matter? A study of consumer product reviews. Journal of Business Research, 63(12), 1336-1341.